\documentclass[prd,onecolumn,nofootinbib,showpacs,showkeys]{revtex4}  
\usepackage{latexsym,graphicx,amssymb,amsmath,mathrsfs} 
\usepackage{setspace,bm} 

\newcommand{\be}{\begin{equation}}      
\newcommand{\ee}{\end{equation}}      
\newcommand{\bea}{\begin{eqnarray}}      
\newcommand{\eea}{\end{eqnarray}}    
     
\newcommand{\rt}[1]{{}}

\makeatletter
\renewcommand\appendix{\par
\setcounter{section}{0}%
\setcounter{subsection}{0}%
\gdef\thesection{\appendixname\space\@Alph\c@section}}
\makeatother

\begin{document}           
{\allowdisplaybreaks   

\title{Spontaneously broken ground states    
of the $U(n)_L\times U(n)_R$ linear sigma model at large $n$}

\author{G. Fej\H{o}s}
\email{geg@ludens.elte.hu}
\affiliation{Department of Atomic Physics, E{\"o}tv{\"o}s University,
H-1117 Budapest, Hungary}
\author{A. Patk{\'o}s}
\email{patkos@galaxy.elte.hu}
\affiliation{Department of Atomic Physics, E{\"o}tv{\"o}s University,
H-1117 Budapest, Hungary}
\affiliation{Research Group for Statistical and Biological Physics 
of the Hungarian Academy of Sciences, H-1117 Budapest, Hungary}

\begin{abstract}        
{Symmetry breaking patterns of the $U(n)_L\times U(n)_R$ symmetric meson model are investigated in a formulation involving three auxiliary composite fields. The effective potential is constructed at leading order in the $1/n$ expansion for a condensate belonging to the center of the  $U(n)$ group. A wide region is found in the coupling space where in addition to the condensate proportional to the unit matrix, also metastable minima exist even in the chiral limit, in which a further breakdown of the diagonal $U_V(n)$ symmetry to $U_V(n-1)$ is realized. Application of a moderate external field conjugate to this component of the order parameter changes this state into the true ground state of  the system. 
}        
\end{abstract}       
\maketitle           
\section{Introduction}  

The numerical details of the pattern of chiral symmetry breaking in theories of three-flavored strong interactions are essentially dictated by the observed mesonic spectra \cite{haymaker74,lenaghan00,black02,tornqvist99}. Potentials of effective models are constructed with the requirement that, in the ground state, a condensate proportional  
to the $3\times 3$ unit matrix should arise spontaneously and split the degeneracy of the parity partner states. This fundamental structure is slightly modified by an external source also proportional to the unit matrix, which generates mass for the pseudo-Goldstone pseudoscalar mesons. The presence of a further external field proportional to the $T_8$ generator results in the kaon-pion mass splitting via the formation of a supplementary condensate along the same direction of the Lie algebra. Finally, the effect of the axial anomaly is invoked for interpreting the emergence of a heavy $\eta$ meson. A complete parametrization of the three-flavored linear sigma model has to rely also on some mass information concerning the scalar sector.

It can be expected that this pattern will not remain unchanged when the number of flavors is increased. 
The investigation of such a sequence of theories is employed when looking for a reliable starting point in constructing the nonperturbative solution of strongly coupled effective meson theories. The solution of the large $n$ extension of the two-flavor theory from an $O(4)$ symmetric theory over an $O(n)$ symmetric one is well-known \cite{zinn02} but does not  bring any new structure in the condensate since only a unique type of vacuum condensate might emerge [up to global $O(n)$ rotations]. Anyhow, this extension found several applications in modeling chiral symmetry breaking in QCD \cite{jako04,brauner08} and also in technicolor theories \cite{appelquist95,kikukawa08}.
Applications to nonequilibrium phenomena, e.g., to the quantum dynamics of disoriented chiral condensates \cite{blaizot92,rajagopal93,bjorken97}, also benefit from this extension.
 
More complex symmetry breaking patterns might arise when the three-flavor $U(3)\times U(3)$ symmetric effective meson theory is generalized to large $n$.
The implementation of the restricted large $n$ extension described above retains only the $O(2n^2=18)$ invariant part in the quartic potential.
Here, the main obstacle in front of wider applications is the missing leading order solution when $n\rightarrow\infty$, which would make explicit the differences between the ground states of the $U(n)\times U(n)$ symmetric and the $O(2n^2)$ symmetric models. A first step in this direction was made with the matrix  generalization of the Hubbard-Stratonovich transformation involving the complete quartic potential of the $U(n)\times U(n)$ symmetric model \cite{frei90}. Based on this rewriting of the model a saddle-point solution was constructed for the field theory defined with the auxiliary matrix field \cite{meyer-ortmanns96}. It was {\it conjectured} that this solution describes the leading order behavior of its ground state for a large flavor number. The approximate solution was applied directly to the thermodynamics of the three-flavor effective meson model (see also \cite{bilic99}). In the actual solution of the matrix saddle-point equation, the authors restricted their Ansatz to a condensate proportional to the unit matrix, which reduced once again the quartic potential to its $O(2n^2)$ invariant part. As a result the part of the vacuum condensate pointing in the algebra space along the $T_8$ direction remained proportional to the $T_8$ component of the explicit symmetry breaking external source. 
 To our actual knowledge no published further step was attempted towards the exact large $n$ analysis of the $U(n)\times U(n)$ symmetric matrix model until the proposition \cite{fejos10} appeared. 

The new feature of \cite{fejos10} was the introduction of two auxiliary fields: beyond an $O(2n^2)$ invariant which corresponds to earlier Ans\"atze, also a $U(n)$ vector 
appeared with nonzero saddle-point value. Following the introduction of these auxiliary fields, the large $n$ limits of the one-point and two-point equations were found and solved with a physically arguable supplementary restriction. 
Although only a condensate proportional to the unit matrix ($v=v_0\times I$) was assumed, the introduction of the new composite field with nonzero saddle value led to new features not present in the large $n$ solution of the $O(2n^2)$ model (for instance, nondegenerate spectra emerged in the scalar meson sector). The supplementary assumption invoked self-consistently during the construction was the heaviness of the scalar fields relative to the
pseudoscalar degrees of freedom. A large domain was found in the three-dimensional coupling space formed by the two independent quartic couplings and the external source $h=h_0\times I$, where this self-consistency is fulfilled.

The present paper is a continuation of \cite{fejos10} and expands the discussion of the proposed solution in three new directions, shortly described below.

A natural next step towards the phenomenological application of this solution is the extension of the condensate by a term proportional to the "longest" diagonal generator: $T_n^{diag}\sim\textrm {diag}(1,1,...,1,-n+1)$, which coincides, for $n=3$, with $T_8$. There is the possibility for introducing a more general condensate belonging to the Cartan subalgebra of $U(n)$, but the more restricted Ansatz $T_0v_0+T_n^{diag}v_n^{diag}$ is the one smoothly connected to the relevant $n=3$ case. In the present study, we construct an approximate large $n$ effective potential at leading order for such a condensate and search for its global minimum with and without the presence of the corresponding explicit symmetry breaking source, $T_n^{diag}h_n^{diag}$.

The second new feature is the replacement of the single composite $U(n)$ vector introduced in \cite{fejos10} by the sum of two fields, each of definite parity. One of them mixes with the scalar, the other with the pseudoscalar elementary fields. The composite pseudoscalar is assumed to be similarly light, like the elementary $\pi^a$; therefore, its dynamics will be included into our analysis, while, in \cite{fejos10}, the sum of the two fields dominated by the scalar composite was classified to be ``heavy,'' and its fluctuations were omitted from the contributions to the equation of state and the saddle-point equations. In the present refined analysis, the mixing in the pseudoscalar sector results in a spectrum, where also heavier excitations ("kaons") might appear naturally. The self-consistency of this solution requires that even the heavier pseudoscalars should be considerably lighter than the lightest scalar excitation. We call the part of the coupling space where this condition is fulfilled the ``allowed space.''
 
It will be shown that, in the allowed coupling space, besides the ``trivial'' solution (e.g., $v_0\neq 0, v_n^{diag}=0$), also a nontrivial local minimum of the effective potential with $v_0\neq0, v_n^{diag}\neq 0$ exists even for $h_0=h_n^{diag}=0$. We shall explore the behavior of this minimum also for $h_0\neq 0$ and find that there is an upper bounding value, beyond which the nontrivial local minimum disappears. 

In order to decide which is the global minimum (i.e., the ground state), we shall map out the effective potential along the $v_n^{diag}$ axis. The construction of this potential and the proof of its renormalizability is the third new result of the present investigation. First, we interpret the behavior of the potential for $h_0\neq 0, h_n^{diag}=0$. We find that the trivial solution is favored energetically everywhere in this subspace. In this way, we demonstrate the global stability of the solution presented in \cite{fejos10} in the absence of explicit symmetry breaking along the $T_n^{diag}$ direction. In some points of the coupling space, where the difference in the potential energy compared to the trivial minimum is relatively small, also the effect of a nonzero $h_n^{diag}$ is explored. Dialing this external source one can tune the state with $v_n^{diag}\neq 0$ to become the true ground state of the model.

In Sec. II, we shortly reformulate the model with three auxiliary fields. In Sec. III the renormalized spectra of the pseudoscalar sector are worked out, since in view of the assumed mass hierarchy, only the fluctuations of these fields contribute to the equations of state and the saddle-point equations. The effective potential is renormalized and evaluated in Sec. IV with the pseudoscalar propagators determined in Sec. III. The numerical solution covering the part of the coupling space where the assumed mass hierarchy is self-consistently valid is presented and discussed in Sec. V. Our conclusions will be summarized in Sec. VI with some hints towards the phenomenological consequences of the new type of ground state described in the paper.

\section{Formulation of the model with auxiliary fields}

The Lagrangian of the model can be written with the help of
$n^2$ pseudoscalar ($\pi^a$) and $n^2$ scalar ($s^a$) components as              
\bea        
L&=&\frac{1}{2}[(\partial_\mu        
s^a)^2+(\partial_\mu\pi^a)^2-m^2((s^a)^2+(\pi^a)^2)]+\sqrt{2n^2}h_as^a        
\nonumber\\        
&-&\frac{g_1}{4n^2}\Big((s^a)^2+(\pi^a)^2\Big)^2-\frac{g_2}{2n}[(U_1^a)^2+(U_2^a)^2],        
\eea        
where the $U(n)$ vectors $U^a_i, i=1,2$, are defined as        
\be        
U_1^a=\frac{1}{2}d^{abc}(s^bs^c+\pi^b\pi^c),\qquad U_2^a=f^{abc}s^b\pi^c,       
\ee        
with $d^{abc}$ and $f^{abc}$ being the symmetric and antisymmetric
structure constants of the $U(n)$ group, respectively. We note that, in \cite{fejos10}, the sum $U_1+U_2$ was treated as a unique
combination, and only two auxiliary fields were introduced. In view of $U_1^aU_2^a=0$, the square of the sum becomes the sum of the two squares as it has already been applied above.
  Correspondingly three auxiliary composite fields are introduced by adding the following constraints to the Lagrangian:        
\be        
\Delta        
L=-\frac{1}{2}\left(X-i\sqrt{\frac{g_1}{2n^2}}((s^a)^2+(\pi^a)^2)        
\right)^2-\frac{1}{2}\left(Y^a_1-i\sqrt{\frac{g_2}{n}}U^a_1\right)^2-\frac{1}{2}\left(Y^a_2-i\sqrt{\frac{g_2}{n}}U^a_2\right)^2.        
\ee 
      
In the sum ${\cal L}\equiv L+\Delta L$, a symmetry breaking pattern corresponding to the shifts    
\bea
&        
s^a\rightarrow s^a+\sqrt{2n^2}v_q\delta^{qa},\qquad U_2^a\rightarrow U^a_2+n\sqrt{2}v_qf_{aqc}\pi^c,\nonumber\\
&
U^a_1\rightarrow        
U^a_1+n\sqrt{2}d_{abq}s^bv_q+n^2d_{apq}v_pv_q.        
\eea
is realized.  The indices $"q"$ denote the diagonal generators of $U(n)$. The index $q=0$ refers to the unit matrix, and $q=2,...,n$ refer to the elements of the Cartan subalgebra of $SU(n)$.   
      
After the introduction of the auxiliary field variables and        
the shifts, the full Lagrangian has the following form:        
\bea        
{\cal L}&=&        
\frac{1}{2}\Big[(\partial_\mu s^a)^2+(\partial_\mu\pi^a)^2-m^2\Big(2n^2v_q^2
+  2\sqrt{2n^2}v_qs^q+(s^a)^2+(\pi^a)^2\Big)\Big]-\frac{1}{2}\left(X^2+(Y^a_1)^2+(Y_2^a)^2\right)\nonumber\\
&+&\sqrt{2n^2}h_q(s^q+        
\sqrt{2n^2}v_q)
+i\sqrt{\frac{g_1}{2n^2}}X\Big(2n^2v_q^2        
+2\sqrt{2n^2}v_qs^q+(s^a)^2+(\pi^a)^2\Big)\nonumber\\        
&+&i\sqrt{\frac{g_2}{n}}\left[Y^a_1U^a_1+n\sqrt{2}v_qd_{abq}Y_1^as^b
+Y_1^an^2d_{apq}v_qv_p\right]
+i\sqrt{\frac{g_2}{n}}\left[Y_2^aU_2^a+n\sqrt{2}v_qf_{aqc}Y_2^a\pi^c\right].        
\label{Eq:shifted-aux-L}        
\eea              
The background $v_q$ introduces mixing of the 
pair $s^a,Y_1^a$ for every value of the index $a$. In addition,
it induces nonzero values for $Y_1^q$, and also a $(2n+1)$-dimensional mixing sector is being formed for the propagators of these degrees of freedom: $s^q,X,$ and $Y_1^q$.
On the other hand, pions will mix (for some index values at least) with the composite field $Y_2^a$, which has negative parity. Since the ground state of strong
interactions respects parity, we expect to have a quantum solution of the model,
where $Y_2^a=0$. Note that, if $Y_2^a$ had nonzero value, scalar and pseudoscalar fields would mix, and the analysis of the mass spectrum would be a lot more complicated. Instead of going through the most general analysis, we assume from the start that $\pi^a$ and $s^a$ fields do not mix and, below, we show that this is consistent
with $Y_2^a=0$.
 
We construct correspondingly a quantum solution, where the saddle-point        
values of $X, Y_1^q$ are nonzero, $Y_2^a=0$, and the two-point functions corresponding to the above-mentioned coupled (mixing) sectors are assumed not to vanish.
In our approximation, only the quantum fluctuations of the light $\pi$-fields enter the equations of state of the condensate and the saddle-point equations; therefore, we 
fully avoid, in this paper the analysis of the effect of the more complex symmetry breaking pattern on the $s$ sector.  The $\pi$ sector breaks up into two separate groups of fields. Those for which $f_{uqw}\neq 0$ form 
 two-dimensional subsectors with the corresponding fields $[Y_2^w,\pi^u]$. Their masses depend on the actual background configuration explicitly. The masses of the pions which do not mix with $Y_2$ (because $f_{uqw}=0$) depend only on the saddle-point values of the auxiliary fields.         

 In the following, we focus our attention on the case where the condensate consists of two components, e.g., $q=\{0,n\}$. Even though this paper is about a strict large $n$ study and, therefore, no specific reference for the $n=3$ case will be given, for the sake of brevity, the notations $T_n^{diag}\equiv T_8$, $v_n^{diag}=v_8$, and $h_n^{diag}=\bar{h}_8$ will be used.

\section{Equations of state, saddle-point equations, and pseudoscalar propagators}        
      
Standard rules \cite{rivers} 
of constructing the derivatives of the quantum effective        
action $\Gamma$ with respect to the fields can be readily applied to (\ref{Eq:shifted-aux-L}).
The contribution involving propagators of the heavy scalars are simply omitted in view of the assumed mass hierarchy.       

In the equations of state (EoS) and the saddle-point equations (SPEs), the useful combination $M^2$ can be introduced:        
\be        
M^2=m^2-\frac{i}{n}(\sqrt{2g_1}X+\sqrt{2g_2}Y_1^0).        
\label{Eq:mass-gap}        
\ee        
The SPEs 
of $Y_1^0, Y_1^8$, and $X$ and the EoS of $v_0$ and $v_8$ take the following form:
\bea
Y_1^0&=&i\frac{\sqrt{2g_2}}{n}\left[n^2(v_0^2+v_8^2)+\frac{1}{2}\int_pG_{\pi^a\pi^a}(p)
\right], \qquad \sqrt{2g_2}X=\sqrt{2g_1}Y_1^0,\nonumber\\
Y_1^8&=&i\sqrt{\frac{g_2}{n}}\left[2n\sqrt{2n}v_0v_8+n^2d_{888}v_8v_8+
\frac{1}{2}d_{8ab}\int_pG_{\pi^a\pi^b}(p)\right],\nonumber\\
0&=&n\sqrt{2}(h_0-M^2v_0)+i2\sqrt{g_2}Y_1^8v_8,\nonumber\\
0&=&n\sqrt{2}(\bar{h}_8-M^2v_8)+i2\sqrt{g_2}Y_1^8v_0+i\sqrt{2g_2n}d_{888}Y_1^8v_8-i\sqrt{\frac{g_2}{n}}f_{a8b}\int_pG_{\pi^a Y_2^b}(p).
\label{Eq:eos-spe}
\eea
Here, propagators $G_{\pi^a \pi^b}$ and $G_{\pi^a Y_2^b}$ refer to the elements of
the complete $2n^2 \times 2n^2$ propagator matrix in the $\pi-Y_2$ sector. The
structure of this matrix (e.g., the mixing of the fields) will be made explicit later in this section.
From the above equations, it is clear that the mass splitting arising in the pion sector plays an important role when one evaluates the tadpole-type contributions. As for the saddle-point equation of the auxiliary field $Y_2$, we obtain $Y_2^a=0$, since the only contributing quantum fluctuations would come from the tadpole of the scalar-pseudoscalar mixing propagator $G_{s^a \pi^b}$, which was previously excluded from the solution on the basis of the parity conservation of strong interactions.  This shows that the $Y_2^a=0$ assumption is consistent.

For the various two-point functions of the $\pi$ sector, it is convenient to introduce the following inverse propagator
commonly appearing in several two-point functions:        
\be        
iD_0^{-1}(p)=p^2-M^2.        
\ee
We substitute for all three-point functions their classical 
expressions, which can be read off from ({\ref{Eq:shifted-aux-L}})
 (and listed explicitly in \cite{fejos10}),
and close the coupled set of Dyson-Schwinger equations at the two-point level (Bare Vertex Approximation, i.e. BVA).  
However, it should be clear that this solution nontrivially differs from the application of BVA to the original formulation without auxiliary fields. 
      
First, we discuss the potential mixing of the $\pi^a$ fields among themselves. Using BVA at the level of the auxiliary field-meson-meson vertices 
and omitting the bubble contribution fully, since at least one heavy scalar line is involved, one finds for the inverse propagator matrix element of two pion fields
\be 
i{\cal G}_{\pi^w \pi^u}^{-1}= iD_0^{-1}\delta_{wu}+i\sqrt{\frac{g_2}{n}}d_{w8u}Y_1^8.
\ee

The only mixing connects the components $0$ and $8$, which scales, however, for $n\rightarrow\infty$ as $1/n$ and is, therefore, neglected in what follows. The mass of all diagonal components stays with $M^2$, except the field "$8$":
\be
M_{88}^2=M^2+i\sqrt{\frac{2g_2}{n}}Y_1^8.
\ee

A large set of $\pi$ fields that consists of coefficients of the nondiagonal generators might mix with the composite fields $Y_2^a$, since
\be
i{\cal G}_{\pi^u Y_2^w}^{-1}=-i\sqrt{2ng_2}f_{u8w}v_8.
\ee
The structure constants $f_{u8w}$ are nonzero only for off-diagonal $T^u,T^w$ generators when $[T^u,T^w]$ is diagonal. Since, in the off-diagonal class, the generators are proportional to the "transmutations" of the Pauli matrices $\sigma_x, \sigma_y$ with nonzero
matrix elements shifted to the row-column positions $(j,k)$, one might refer
to them using the extended index notation $(x,j,k)$, $(y,j,k)$. One quickly recognizes that each of the $2(n-1)$ pairs $u=(x,j<n,n), w=(y,j<n,n)$ and $u=(y,j<n,n),w=(x,j<n,n)$ form coupled $2\times 2$ sectors of $\pi^u,Y_2^w$. In the limit $n\rightarrow\infty$, one has $f_{(x,j,n),(y,j,n),8}=-1/\sqrt{2}+{\cal{O}}(1/n)$. Below, wherever $\pi$ fields with these index values $u,w$ occur, we simply use the index $K$ with reference to kaons. Also note that, starting from the definition of the structure constants, one finds that the only nonzero $d$ coefficients at $n=\infty$ involving off-diagonal generators are $d_{(x,j,n),(x,j,n),8}=d_{(y,j,n),(y,j,n),8}=-1/\sqrt{2}+{\cal{O}}(1/n)$. 

The other off-diagonal $\pi$ fields do not mix and stay with the mass parameter $M^2$. Recalling the $"88"$ component with mass $M_{88}^2\neq M^2$, one has $n^2-2(n-1)-1=(n-1)^2$ pseudoscalar fields staying with mass $M^2$, which corresponds to the breakdown of the
diagonal symmetry $U(n)$ to $U(n-1)$. Although the mass corrections of these "pions" are suppressed in the limit $n\rightarrow\infty$, we write them explicitly, since they are essential for the renormalization of Eqs. (\ref{Eq:eos-spe}):
\be
M_\pi^2=M^2-i\frac{\sqrt{2g_2}}{n\sqrt{n}}Y_1^8\equiv M^2+\frac{1}{n}\Delta M_\pi^2.
\label{Eq:pi-mass}
\ee

The $2\times 2$ coupled two-point matrix of the $(K,Y_2)$ sector looks at $n=\infty$ as follows:  
\be
\displaystyle
i{\cal G}_{(K, Y_2)}^{-1}=
\begin{pmatrix} 
~iD_0^{-1}-i\sqrt{\frac{g_2}{2n}}Y_1^8~  &        
~-i\sqrt{ng_2}v_8~\\        
~-i\sqrt{ng_2}v_8~ &~-1~        \end{pmatrix}.      
\label{Eq:pi-y2-sector}        
\ee   
Note that, in (\ref{Eq:pi-y2-sector}), the $K$ index refers to a $(x,j,k)$ type of generator, while the corresponding $Y_2$ carries the appropriate $(y,j,k)$ type of index. There are also $(n-1)$ identical pairs with these indices reversed, where the off-diagonal elements of the $2\times 2$ matrix receive an additional minus sign due to the totally antisymmetric nature of the $f^{abc}$ structure constants.
The corresponding "kaon" mass is given by the zero of its determinant:
\be
M_{K}^2=M^2+i\sqrt{\frac{g_2}{2n}}Y_1^8+ng_2v_8^2.
\ee

The specific interest of the $(K,Y_2)$ propagator matrix (\ref{Eq:pi-y2-sector}) is that it requires a nontrivial
 large $n$ scaling for $Y_1^8$ and $v_8$, if one wants to arrange finite contributions to $M_{K}^2$:
\be
v_8=\frac{\chi_8}{\sqrt{n}},\qquad Y_1^8=\sqrt{n}y_1^8.
\label{Eq:scaling}
\ee

It turns out that also the external source $\bar{h}_8$ should scale as 
$\bar{h}_8=h_8/\sqrt{n}$. 
As a consequence, the $v_8$ condensate decouples from the EoS of $v_0$ and the SPEs of $Y_1^0$ and $X$:
\be
0=n\sqrt{2}(h_0-M^2v_0), \qquad y_1^0=i\sqrt{2g_2}\left[v_0^2+\frac{1}{2}T_F(M^2)\right],\qquad \sqrt{g_2}x=\sqrt{g_1}y_1^0.
\label{Eq:n-squared}
\ee
Here the notations $y_1^0=Y_1^0/n$, $x=X/n$ are introduced, and $T_F(M^2)$ is the finite part of the tadpole integral of mass $M$:
$T_F(M^2)=i\int_k(k^2-M^2)^{-1}-T_d^{(2)}-(M^2-M_0^2)T_d^{(0)}$, where $M_0$ plays the role of the renormalization scale. The definitions of the quadratically divergent $T_d^{(2)}$ and the logarithmically divergent $T_d^{(0)}$ quantities were given in \cite{fejos10}. Their explicit form defines the renormalization scheme we used in the numerical part of our investigation. Here and below, we simply omit the divergent pieces from the tadpole integrals appearing in the EoS and SPEs. In the next section, where the renormalized effective potential is presented, the correctness of this formal procedure is demonstrated.

When one makes use of the off-diagonal element of the $2\times 2$ kaon-propagator ${\cal{G}}_{(K,Y_2)}$, calculated as the inverse of (\ref{Eq:pi-y2-sector}), one arrives, for the EoS (using the large $n$ limit of the relevant structure 
constants) at
\be
0=\sqrt{2n}\left(h_8-M^2\chi_8+i\sqrt{2g_2}y_1^8(v_0-\chi_8)-g_2\chi_8T_F(M^2_K)\right).
\label{Eq:eos-8}
\ee
The  corresponding saddle-point equation looks also quite simple, after one recognizes that $d_{8aa}T_F(M^2)=0$ by $d_{8aa}=0$. Therefore, one adds and subtracts from the contributions of the kaon-tadpoles the pion tadpoles and finds
\be
y_1^8=i\sqrt{2g_2}\left[2v_0\chi_8-\chi_8^2-\frac{1}{2}T_F(M_K^2)+\frac{1}{2}T_F(M^2)\right].\label{Eq:spe-8}
\ee
One can solve the equations of the 8 components, $\chi_8,y_1^8$, with $v_0, y_1^0, x$ taken from their respective Eqs. (\ref{Eq:n-squared}), solved for a set of $h_0, g_1$, and $g_2$. 
These latter part of the calculation coincides with the solution presented in \cite{fejos10}, where only the $v_0$ condensate was considered.

\section{The renormalized effective potential}

In case one finds nontrivial solutions for $\chi_8, y_1^8$ also when $h_8=0$ beyond the trivial $\chi_8=y_1^8=0$ solution,
 one has to decide which solution is preferred energetically, e.g., one has to evaluate the effective potential in the "light $\pi$" approximation described above. This means that, beyond the classical potential, written with the help of the auxiliary variables, also the contribution from the quantum fluctuations of the light particles, e.g., of the pions and kaons, should be included. The corresponding formal expression is the following, with ${\cal O}(n)$ accuracy:
\bea
V&=&V_{cl}+V_{quant},\nonumber\\
V_{cl}&=&n^2\left[M^2v_0^2+\frac{1}{2}(x^2+(y_1^0)^2)-2h_0v_0\right]+n\left[M^2\chi_8^2+\frac{1}{2}(y_1^8)^2-2 h_8\chi_8-i\sqrt{2g_2}y_1^8\chi_8(2v_0-\chi_8)\right],
\nonumber\\V_{quant}&=&-\frac{i}{2}\left[(n^2-2n)\int_p \ln(-p^2+M_\pi^2)+2n\int_p\ln(-p^2+M_K^2)\right],
\label{Eq:eff-pot}
\eea
where $V_{quant}$ is the standard expression of the one-loop part of the effective potential.
It is important to note that, for the accurate evaluation of the quantum contribution of the pions, one has to make use of the ${\cal O}(n)$ accurate value of $M_\pi^2$, as given in (\ref{Eq:pi-mass}).

The divergences of $V_{quant}$ can be eliminated by subtracting appropriate constants with divergences separated in terms of the auxiliary propagator $i/(p^2-M_0^2)$ in a way to have compatibility with the previously obtained finite Eqs. (\ref{Eq:n-squared}), (\ref{Eq:eos-8}) and (\ref{Eq:spe-8}). The counterterm part of the potential which absorbs these divergencies is the following:
\bea
V^{ct}&=& -\frac{n^2}{2}\left[(M^2-m^2)(T_d^{(2)}-M_0^2T_d^{(0)})+\frac{1}{2}(M^4-m^4)T_d^{(0)}\right]+i\frac{n^2}{2}\int_p\ln (-p^2+M_0^2)\nonumber\\
&&-n\left[\left(\frac{1}{2}\Delta
M_\pi^2+M_K^2-M^2\right)(T_d^{(2)}+(M^2-M_0^2)T_d^{(0)})+\frac{1}{2}(M^2_K-M^2)^2T_d^{(0)}\right],
\label{Eq:ct}
\eea
Although (\ref{Eq:ct}}) is written in a rather compact form, it is a function of $v_0,\chi_8,y_1^0,y_1^8,$ and $x$, and is completely allowed by the formal requirements of renormalizability \cite{berges05}.
Note also the important simplification $\Delta M_\pi^2/2+M_K^2-M^2=g_2\chi_8^2$. One can check explicitly that
the derivatives of the renormalized potential (\ref{Eq:eff-pot}) below, with respect to $v_0,\chi_8,y_1^0,y_1^8$ and $x$, reproduce the finite
Eqs. (\ref{Eq:n-squared}),(\ref{Eq:eos-8}) and (\ref{Eq:spe-8}). One should also
recognize that the second term of the right-hand side of the first line of (\ref{Eq:ct}) is field independent, and,
therefore, gives no contribution to the equations for the one-point functions. It serves
only for eliminating the overall divergence generated by the zero-point fluctuations of the pions.

The quantum part of the renormalized effective potential ($V_{quant,R}\equiv V_{quant}+V^{ct}$) in the renormalization scheme outlined above looks like
\bea
V_{quant,R}&=&\frac{n^2}{32\pi^2}\left[\frac{3}{4}(M_0^4-M^4)+M_0^2(M^2-M_0^2)+\frac{1}{2}M^4\ln\frac{M^2}{M_0^2}\right]\nonumber\\
&+&\frac{n}{16\pi^2}\left[\frac{1}{2}(\Delta M_\pi^2-M^2)M^2\ln\frac{M^2}{M_0^2}
+\frac{1}{2}M_K^4\ln\frac{M_K^2}{M_0^2}+g_2\chi_8^2M_0^2-\frac{1}{2}\Delta M_\pi^2M^2+\frac{3}{4}(M^4-M_K^4)\right].
\label{Eq:eff-pot}
\eea
At this point, we have the complete renormalized effective potential, $V_R\equiv V_{cl}+V_{quant,R}$, and we can decide which solution is more favorable energetically.
Since we found two solutions for $\chi_8, y_1^8$ in a fixed background of $v_0,x,y$
values, we are only looking for the $\chi_8, y_1^8$ dependent parts of the effective
potential to select the true ground state of the system. One can immediately see that
only the ${\cal{O}}(n)$ part of $V_R$ is what we have to study. It is also important to
stress that, due to this fortunate effect, the selection of the ground state does not
depend on the explicit value of $n$, since it appears as an overall multiplicative factor
of every relevant term. In the forthcoming figures, we refer to $V^{(n)}$ as the
${\cal{O}}(n)$ part of the full effective potential without the overall $n$ factor.

The fact that the $8$ condensate contributes only to ${\cal O}(n)$ 
means that the $v_8$ contribution will be subleading relative to that of $v_0$. 
Still, it belongs to the leading large $n$ piece, since the next-to-leading-order contributions which were omitted from the very start would contribute ${\cal O}(n^0)$.

\section{Numerical results}
 
In \cite{fejos10}, we mapped out the region of the three-dimensional parameter space $(g_1,g_2,h_0)$, where the scalar masses calculated from the solution of 
(\ref{Eq:n-squared}) turned out to be at least twice as large as the mass of the pseudo-Goldstone fields (pions). In the present investigation, we have chosen rather low values 
of $h_0$ (all dimensional quantities are measured in proportion to the appropriate powers of the absolute value of the renormalized mass). Note, however, that, in this paper, we have chosen $M_0^2=-m^2$ for the renormalization scale, different from the choice of \cite{fejos10};
 therefore Fig.2 of that paper can not be directly compared to this study. We just state that the results below are obtained in those regions of the parameter space, where the heavy scalar assumption holds (i.e., where the scalar masses are at least twice as large as the pion mass). This assumption will be checked if it also holds for the scalar/kaon mass ratio as well.

\begin{figure}
\includegraphics[bb=485 90 340 380,scale=0.63]{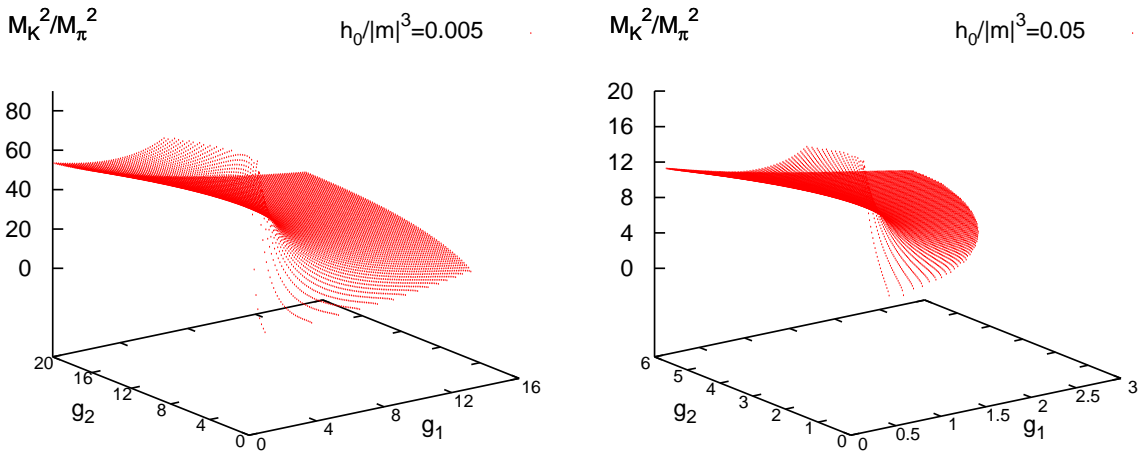}
\caption{Kaon/pion squared mass ratio as a function of the couplings ($g_1,g_2$) for two different $h_0$ values. Kaons become lighter as $h_0$ increases, at the same time the region where nontrivial solutions are found narrows.}
\label{Fig1}
\end{figure}

Let us start the discussion with the mass splitting in the $\pi$ sector - in particular, the kaon/pion mass ratio. 
The most detailed study was achieved for $h_8=0$, addressing the question of the $v_8$ (e.g., $\chi_8$) dependence of the effective potential after eliminating the $y_1^8$ dependence of $V_R$ with the help of (\ref{Eq:spe-8}). We scanned through the allowed region at different values of $h_0$. Our main result is that, in addition to the trivial solution $v_8=iy_1^8=0$, also a nontrivial solution of 
(\ref{Eq:eos-8}) and (\ref{Eq:spe-8}) with positive squared masses of all $\pi$ fields was found in a large
part of the region allowed by the heavy scalar mass assumption. In Fig.\ref{Fig1}, the ratio $M_K^2/M_\pi^2$ is displayed over the $(g_1,g_2)$ plane for 2 different
values of $h_0$. One can see that, in a quite large region of the allowed parameter space, the mass ratio varies mildly, but the kaons are getting
heavier when $h_0$ is lowered. Also, it can be observed that, for small $g_1$ values, increasing $g_2$ induces very large kaon mass values, while, for larger $g_1$, the same does not lead to significant changes of the mass ratio. 

\begin{figure}
\includegraphics[bb=545 50 340 580,scale=0.4]{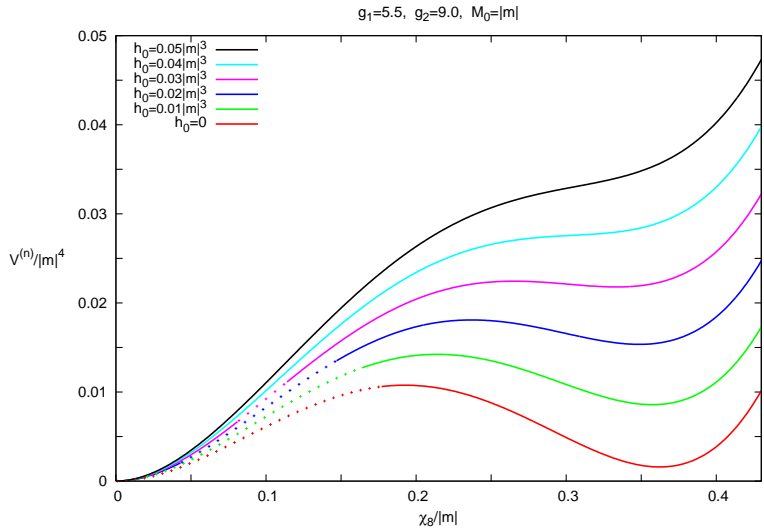}
\caption{The ${\cal{O}}(n)$ part of the quantum effective potential divided by $n$ as a function of the $\chi_8$ condensate. The real part of the potential is plotted with dotted lines for $\chi_8$ values where some negative mass squares lead to complex values of the potential. The curves from top to bottom follow the order of the $h_0$ values indicated in the upper-left-hand corner.}
\label{Fig2}
\end{figure}

The ${\cal O}(n)$ part of the full renormalized effective potential $V_R$ could be plotted in an extended interval around the nontrivial minimum, where all mass squares are positive. 
An example is shown in Fig.\ref{Fig2}, where one can see that the energy density of the nontrivial solution grows monotonically with $h_0$ increasing. [As already announced, the ${\cal{O}}(n)$ part of the full potential divided by $n$ is denoted as $V^{(n)}$, and the same holds for the classical part as $V_{cl}^{(n)}$.] There is some critical
value $h_0^c(g_1,g_2)$, where the nontrivial local minimum disappears. For example, one can verify on Fig. \ref{Fig1} that, for the $g_1,g_2$ values of Fig. \ref{Fig2}, there does exist a nontrivial solution of the equations for $h_0=0.005$ but not for $h_0=0.05$. 

The dotted pieces of the curves represent the real part of the effective potential in regions where it becomes complex due to the fact that one of the mass squares becomes negative. This phenomenon is a well-known feature of the loop expansion \cite{rivers,fujimoto83,raifear86}. Although the effective potential should be a real quantity by construction, in regions where the classical potential is chosen to be nonconvex, the loop expansion breaks down and produces an imaginary part. Mathematically, a nonconvex classical potential can be understood as an analytic continuation of a convex classical potential through the mass term:
$m \rightarrow im$. The loop expansion creates a divergent asymptotic series, but there is no such theorem which would state that the analytically continued series corresponds term-by-term to the series of the analytically continued effective potential. In other words, the analytic continuation and the expansion in $\hbar$ are not interchangable operations. From a mathematical point of view, this is the origin of the appearance of the imaginary part. Nevertheless, the naive calculation leading to complex effective potentials has a physical meaning. First, it turns out that, if one wants to understand the effective potential at a $\phi_c$ point as the expectation value of the minimum energy density of the class of quantum states in which the field expectation value is $\phi_c$ (which property can be derived from the usual Legendre transform definition of the effective potential), then one has to take the {\it real part} of the loop expansion and apply Maxwell's construction. Furthermore, Weinberg and Wu showed \cite{weinberg87} that even the naively calculated real part in itself contains physics: it is the minimum energy density of a class of quantum states with a $\phi_c$ field expectation value, together with the more restrictive constraint that their wave functionals are concentrated on the uniform configuration $\phi_c$. The state chosen this way may differ from the state which {\it minimizes} the energy density of states with a $\phi_c$ field expectation value. In this case, the state in question is unstable, and the {\it imaginary part} of the potential can be interpreted as half the decay rate per unit volume. 

\begin{figure}
\includegraphics[bb=505 50 340 420,scale=0.5]{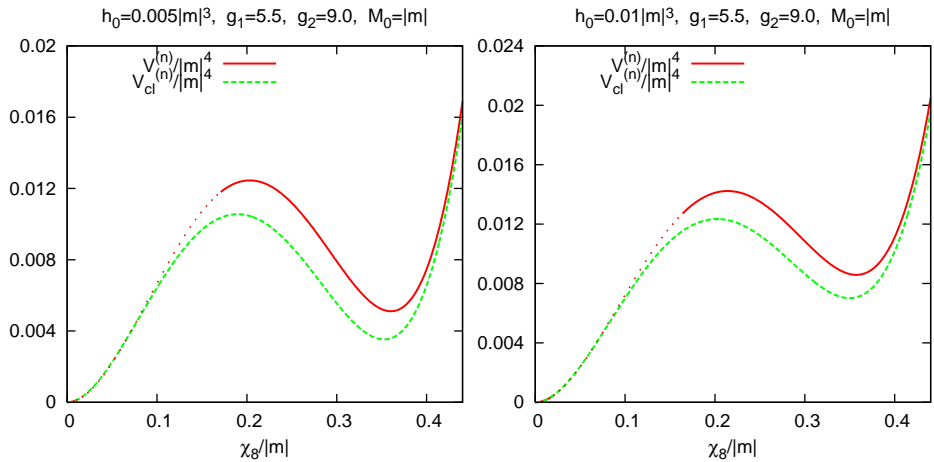}
\caption{The effect of the quantum fluctuations is only moderate, as indicated by comparing the $V_{cl}^{(n)}$  (dashed lines) with $V^{(n)}$. The dotted part displays the real part of $V^{(n)}$, where it becomes complex.}
\label{Fig3}
\end{figure}

We have also plotted the curve
of the classical part of the ${\cal{O}}(n)$ piece of the potential in a typical point of the coupling space for two values of $h_0$, which demonstrates that, with the present renormalization scale $M_0$ the quantum fluctuations only moderately modify the value of the potential; see Fig.\ref{Fig3}.
In the whole allowed region of the coupling space, we found for the potential in the nontrivial minimum positive
values; therefore, we conclude that, with $h_8=0$, there is no $h_0,v_0, M^2$ combination which would induce a spontaneously nonzero $v_8$ condensate into the true ground state of the system.
The effect of the existence of such a metastable state might still show up in out-of-equilibrium situations.

\begin{figure}
\includegraphics[bb=505 50 340 580,scale=0.42]{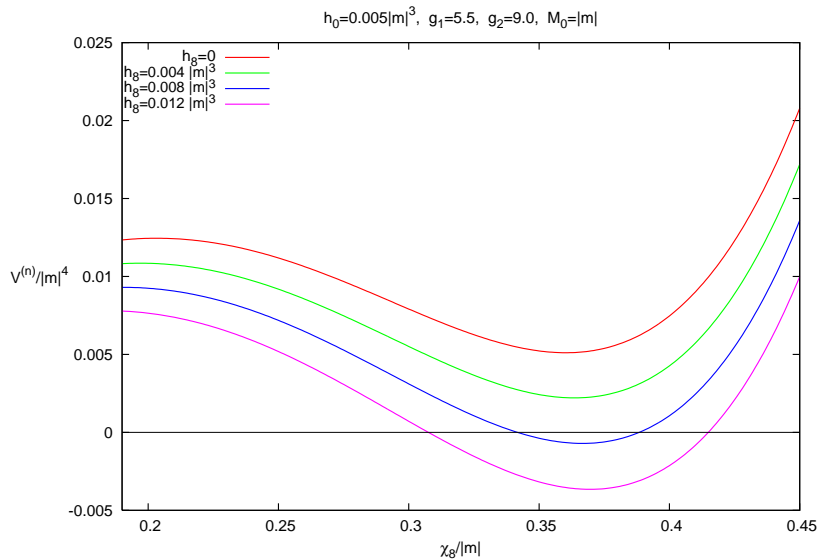}
\caption{Smooth variation of $V^{(n)}$ with increasing $h_8$ displaying the symmetry breaking pattern $U_V(n)\rightarrow U_V(n-1)$. The curves from top to bottom follow the order of the applied $h_8$ values indicated in the upper-left-hand corner.}
\label{Fig4}
\end{figure}

In a number of energetically "promising"
coupling points, we have studied the effect of the application of a nonzero $h_8$ on the effective potential.
In these cases, also the trivial solution shifts in proportion of the external field, and the effective potential can be displayed in a finite but small region
around the shifted trivial minimum, too (i.e., no negative mass squares appear in its neighborhood). If $h_0$ is small, the value of the effective potential in the nontrivial minimum
varies faster with $h_8$ than near the shifted trivial minimum; therefore, one finds a critical external field $h_8^c(g_1,g_2)$ which drives the minimum characterized by a large $v_8$ value to be the true ground
state of the system. In Fig.\ref{Fig4}, we show the typical variation of the potential as a function of $h_8$ for the value of $(g_1,g_2,h_0)$ chosen as in the left panel of Fig.\ref{Fig3}.
\begin{figure}
\includegraphics[bb=445 90 340 580,scale=0.6]{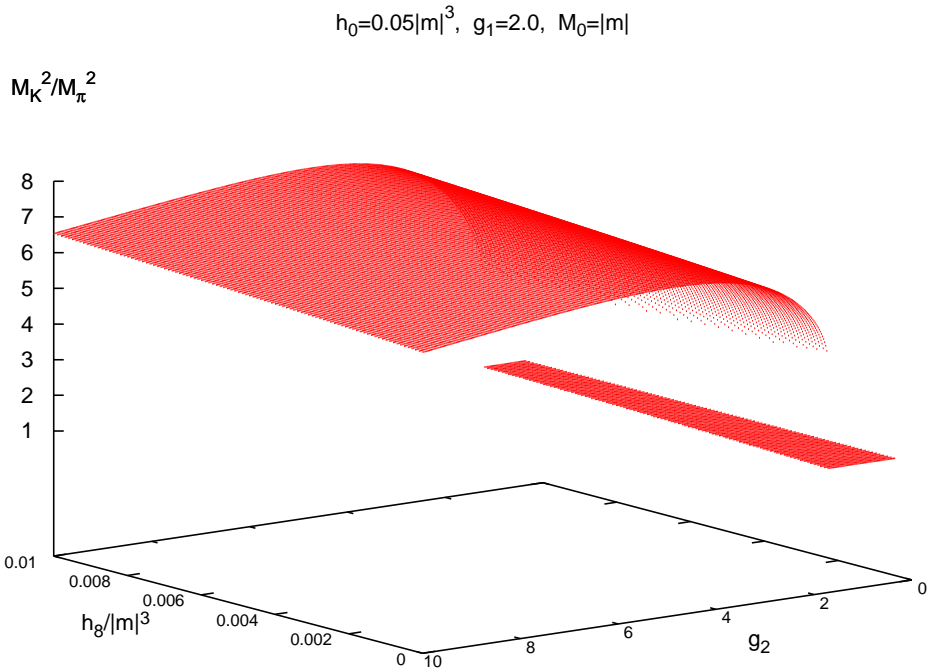}
\caption{The variation of $M_K^2/M_\pi^2$ with $h_0$ and $g_2$. The relative variation of the mass ratio is very small with respect to $h_8$. Note that the
abrupt jump downwards is occurring when the nontrivial minimum is missing.}
\label{Fig5}
\end{figure}
In Fig.\ref{Fig5}, we display the variation of the ratio $M_K^2/M_\pi^2$ as a function of $h_8$ and $g_2$ for given $g_1$ and $h_0$ values as taken in the nontrivial minimum. One can see that,
apart from a jump to unity happening at lower $g_2$ values, the ratio is not varying too much. This ratio equals unity in the region where the nontrivial minimum is missing. Choosing higher values for $h_0$ involves smaller kaon masses (as it can also be seen in Fig.\ref{Fig1}), which points to the direction in the parameter
space with a chance of phenomenological applications. One also has to note that, choosing other $g_1$
values, the shape of the figure changes only mildly. The figure also shows that the mass ratio is increasing linearly with $h_8$ at fixed $g_2$; however, this change is 
almost negligible compared to its full magnitude, i.e., the relative change is very small. This means that the kaon mass is induced dominantly by the action of
the $v_0$ condensate and not by the explicit symmetry breaking proportional to $h_8$. This circumstance is rather different from what comes in the $n=3$ flavor models from various perturbative discussions.

Finally, we mention that it was checked that the scalar masses are also at least twice heavier than the kaons in the overwhelming
part of the allowed parameter space, which means that our approximation based on the specific scalar/pseudoscalar
mass hierarchy remains valid still in a wide range of parameters.

\section{Summary and outlook}

Symmetry breaking patterns emerging in the ground state structure of the $U_L(n)\times U_R(n)$ symmetric matrix model were explored at leading order in the $1/n$ expansion.
For the first time, ground states genuinely different from the broken $O(2n^2)$ symmetry were studied  with a resummed perturbative approach, including the effect of the quantum fluctuations of the light (pseudo-Goldstone) fields. The effective potential of this approximation was consistently renormalized and used to
compare energetically local minima corresponding to different symmetry breaking condensates. It was shown that in a large part of the coupling space,
a condensate breaking $U_L(n)\times U_R(n)$ to $U_V(n)$ induces also the existence of a metastable local minimum with a further symmetry breaking $U_V(n)\rightarrow U_V(n-1)$.  The effect of the explicit symmetry breaking sources on the local minima and the spectra was studied in detail. It was established that the application of an appropriate combination of the external sources drives quite often the $U_L(n)\times U_R(n)\rightarrow U_V(n-1)$ breaking local minimum into the true ground state of the system. 

The different ground states are characterized by different pseudoscalar excitation spectra, also explicitly constructed. The mass ratio of the heavier modes ("kaons") and the
remaining $(n-1)^2$ pseudo-Goldstone modes was found sensitive only to the external source $h_0$ proportional to the $T^0$ generator. The mild dependence on $h_n^{diag}T_n^{diag}$ is rather different from what characterizes the kaon/pion ratio in perturbative treatments of the  $U_L(n)\times U_R(n)\rightarrow U_V(n)$ symmetry breaking. If it turns out that one can build phenomenology on the nontrivial vacuum condensate, this observation would imply also rather different finite temperature behavior, compared to the perturbative descriptions.
 
Although the transition in the $v_0$ condensate in itself would be much similar to that of the $O(2n^2)$ symmetric model, the insensitivity of the location of the nontrivial minimum in $v_n^{diag}$ with respect to $h_n^{diag}$ (see Fig.\ref{Fig5}) hints to a first-order transition. We observed that it is not $h_n^{diag}$ but rather $v_0$ which determines the value of $v_n^{diag}$. The tightly connected variation of the two condensates probably leads then to a phase transition with both fields going through a discontinuous change of similar amplitude. This expectation complies well with renormalization group considerations which classify the restoration of the broken symmetry in this model quite differently than in models displaying symmetries of the orthogonal group \cite{paterson81,pisarski83}. An obvious next step is the study of the finite temperature symmetry restoration from this nontrivial ground state. 

Further consolidation of the discovered structure requires the investigation of the fluctuation effect of the heavy scalar fields. On a general basis, one expects the modification of the different coefficients in different EoS, SPEs, and self-energies by terms of ${\cal O}(M_\pi^2/M_s^2)$ (where $M_s^2$ is one of the scalar masses), which would just deform slightly the coupling region where the nontrivial vacuum structure is self-consistently present.

Our final goal is to make use of this solution at $n=3$, when also the contributions from the effective term reflecting the axial anomaly (the 't Hooft determinant) would be included perturbatively using the large-$n$ propagators of the different fields.The comparison with the treatments of the $n=3$ three flavored meson model in different perturbative calculations \cite{dumm05,dumm06,kovacs08,schaefer09,schaefer10} and functional approaches, which go beyond perturbation theory \cite{braun04,jaeckel04}, partly based on the exact renormalization group techniques \cite{berges02,gies06,schaefer08}, could give some insight into the importance of strong coupling effects. The further exploration of the proposed ground state will proceed along these three lines.

\begin{acknowledgments}
The authors dedicate this paper to Prof. P{\'e}ter Sz{\'e}pfalusy on the occasion of his 80th birthday. 
This work is supported by the Hungarian Research Fund under Contracts
No. T068108 and K77534. 
\end{acknowledgments}


\begin{thebibliography}{9}
\bibitem{haymaker74}L.-H. Chan and R.W. Haymaker, Phys. Rev. D{\bf 7}, 415 (1973)
\bibitem{lenaghan00}J.T. Lenaghan, D.H. Rischke and J. Schaffner-Bielich, Phys. Rev. D{\bf 62}, 085008 (2000)
\bibitem{black02}D. Black, A.H. Fariborz, S. Moussa, S. Nasri and J. Schechter, AIP Conf. Proc. {\bf 619}, 179 (2002)
\bibitem{tornqvist99}N.A. Tornqvist, Eur. Phys. J. C{\bf 11}, 359 (1999)
\bibitem{zinn02}J. Zinn-Justin, {\it Quantum Field Theory and Critical Phenomena} (Clarendon Press, Oxford, 2002) 4th ed.
\bibitem{jako04}A. Jakov\'ac, A. Patk{\'o}s, Zs. Sz{\'e}p and P. Sz{\'e}pfalusy, Phys. Lett. B{\bf 582}, 179 (2004)
\bibitem{brauner08}J.-O. Andersen and T. Brauner, Phys. Rev. D{\bf 78}, 014030 (2008)
\bibitem{appelquist95} T. Appelquist, M. Schwetz and S.B. Selipsky, Phys. Rev. D{\bf 52}, 4741 (1995)
\bibitem{kikukawa08} Y. Kikukawa, M. Kohda and J. Yasuda, Phys. Rev. D{\bf 77}, 015014 (2008)
\bibitem{blaizot92}J.P. Blaizot and A. Krzywicki, Phys. Rev. D{\bf 46}, 246 (1992)
\bibitem{rajagopal93}K. Rajagopal and F. Wilczek, Nucl. Phys. B{\bf 399}, 395 (1993)
\bibitem{bjorken97}J.D. Bjorken, Acta Phys. Pol. B{\bf 28}, 2773 (1997)
\bibitem{frei90}Z. Frei and A. Patk\'os, Phys. Lett. B{\bf 247}, 381 (1990)
\bibitem{meyer-ortmanns96}H. Meyer-Ortmanns and B.J. Schaefer, Phys. Rev. D{\bf 53}, 6586 (1996)
\bibitem{bilic99}N. Bilic and H. Nikolic, Eur. Phys. J. C{\bf 6}, 515 (1999)
\bibitem{fejos10} G. Fej\H{o}s and A. Patk{\'o}s, Phys. Rev. D{\bf 82}, 045011 (2010)
\bibitem{rivers}R.J. Rivers, {\it Path Integral Methods in Quantum Field Theory} (Cambridge University Press, Cambridge, 1987)
\bibitem{berges05}J. Berges, S. Borsanyi, U. Reinosa and J. Serreau, Ann. Phys. {\bf 320}, 344 (2005)
\bibitem{fujimoto83}Y. Fujimoto, L. O'Raifeartaigh, G. Parravicini, Nucl. Phys. B{\bf 212}, 268 (1983)
\bibitem{raifear86}L. O'Raifeartaigh, A. Wipf and H. Yoneyama, Nucl. Phys. B{\bf 271}, 653 (1986)
\bibitem{weinberg87}E.J. Weinberg and A. Wu, Phys. Rev. D{\bf 36}, 2474 (1987)
\bibitem{paterson81}A.J. Paterson, Nucl. Phys. B{\bf 190}, 188 (1981)
\bibitem{pisarski83}R.D. Pisarski and F. Wilczek, Phys. Rev. D{\bf 29}, 338 (1984)
\bibitem{dumm05}D. Gomez Dumm and N.N. Scoccola, Phys. Rev. C{\bf 72}, 014909 (2005)
\bibitem{dumm06}D. Gomez Dumm, A.G. Grunfeld and N.N. Scoccola, Phys. Rev. D{\bf 74}, 054026 (2006)
\bibitem{kovacs08} P. Kov\'acs and Zs. Sz{\'e}p, Phys. Rev. D{\bf 77}, 065016 (2008)
\bibitem{schaefer09} B.J. Schaefer, M. Wagner, Phys. Rev. D{\bf 79}, 014018 (2009)
\bibitem{schaefer10} B.J. Schaefer, M. Wagner and J. Wambach, Phys. Rev. D{\bf 81}, 074013 (2010)
\bibitem{berges02} J. Berges, N. Tetradis and C. Wetterich, Phys. Rept. {\bf 363}, 223 (2002)
\bibitem{gies06} H. Gies, arXiv:0611.146 [hep-ph]
\bibitem{schaefer08} B.J. Schaefer and J. Wambach, Phys. Part. Nucl. {\bf 39}, 1025 (2008)
\bibitem{braun04} J. Braun,  H.J. Pirner and K. Schwenzer, Phys. Rev. D{\bf 70}, 085016 (2004)
\bibitem{jaeckel04} J. Jaeckel and C. Wetterich, Nucl. Phys. A{\bf 733}, 113 (2004) 
\end{thebibliography}
 \end{document}